\documentclass[aps,prl, twocolumn,superscriptaddress,showpacs, longbibliography]{revtex4-1}

\usepackage{amssymb}
\usepackage{amsmath}
\usepackage{graphicx}
\usepackage{natbib}
\usepackage{nicefrac}
\usepackage{multirow}
\usepackage{epstopdf}
\usepackage{float}
\usepackage{physics}

\usepackage[ linktocpage=true, colorlinks, allcolors=blue]{hyperref}

\usepackage{color} 

\begin{document}

\title{Microwave detection of electron-phonon interactions \protect\\ in a cavity-coupled double quantum dot}

\author{T.~R.~Hartke}
\affiliation{Department of Physics, Princeton University, Princeton, New Jersey 08544, USA}
\author{Y.-Y.~Liu}
\affiliation{Department of Physics, Princeton University, Princeton, New Jersey 08544, USA}
\author{M.~J.~Gullans}
\affiliation{Department of Physics, Princeton University, Princeton, New Jersey 08544, USA}
\author{J.~R.~Petta}
\affiliation{Department of Physics, Princeton University, Princeton, New Jersey 08544, USA}


\begin{abstract}
Quantum confinement leads to the formation of discrete electronic states in quantum dots. Here we probe electron-phonon interactions in a suspended InAs nanowire double quantum dot (DQD) that is electric-dipole coupled to a microwave cavity.  We apply a finite bias across the wire to drive a steady state population in the DQD excited state, enabling a direct measurement of the electron-phonon coupling strength at the DQD transition energy. The amplitude and phase response of the cavity field exhibit features that are periodic in the DQD energy level detuning due to the phonon modes of the nanowire. The observed cavity phase shift is consistent with theory that predicts a renormalization of the cavity center frequency by coupling to phonons.
\end{abstract}

\maketitle

Phonons, the quantized lattice vibrations in a crystalline solid, are ubiquitous in condensed matter systems and impact the properties of bulk and nanostructured materials. For example, Raman scattering measurements provide a detailed probe of the phonon density of states in carbon nanotubes \cite{Dresselhaus05_PhysRep} and it is now known that phonons limit the maximum current in single nanotube devices \cite{Yao00_PRL, Javey03_Nature, Park04_NanoLett}. Similarly, spin relaxation in semiconductors is generally governed by processes that involve phonon emission \cite{Khaetskii01_PRB, vanderWiel02_RevModPhys, Hanson07_RevModPhys}. Electronic relaxation processes involving phonons can even be harnessed to cool mechanical degrees of freedom in nanostructures \cite{Wilson-Rae04_PRL, Zippilli09_PRL}. 

Semiconductor double quantum dots (DQDs) are well-suited for probing electron-phonon coupling \cite{Brandes99_PRL, Brandes05_PhysRep, Weber09_PSSB, Weber10_PRL, Roulleau11_NatCommun} since inelastic interdot tunneling generally involves spontaneous emission of a phonon in order to conserve energy \cite{Fujisawa98_Science}. Moreover, the DQD energy level difference is electrically tunable, which allows the effective electron-phonon coupling strength to be probed as a function of phonon energy. Confinement can further influence phonon modes in nanostructures and new methods of controlled nanowire placement \cite{Flohr11_RevSciInstrum, Nilsson08_NanoLett} may allow for careful engineering of the phonon spectrum in semiconductors, similar to recent optomechanics experiments with suspended carbon nanotubes \cite{Aspelmeyer14_RMP, Laird12_NanoLett, Ares16_PRL}.

In this Letter, we investigate the interplay of electrons, phonons, and photons in a cavity-coupled InAs nanowire DQD which is mechanically suspended in vacuum. The interaction of photons and electrons in DQDs has been studied extensively in the circuit QED architecture, where the charge dipole of a DQD is coupled to a microwave cavity \cite{Frey12_PRL, Delbecq11_PRL,Petersson12_Nature}. A phonon sideband has recently been observed in DQD masers that are driven by single electron tunneling \cite{Liu15_Science} and related theory suggests that the detailed energy-dependence of the one-dimensional nanowire phonon spectral density should have observable consequences in the photon emission rate \cite{Gullans15_PRL}, although this has not yet been observed. Here we measure the dc current $I$ as a function of DQD energy level detuning $\epsilon$ and show that it exhibits periodic oscillations that are consistent with the phonon spectral density in a confined nanostructure \cite{Weber10_PRL, Roulleau11_NatCommun}. Measurements of the cavity amplitude and phase response reveal the detailed energy dependence of the electron-phonon coupling and exhibit a response that is periodic in $\epsilon$. We employ a microscopic theoretical model of the device which suggests that the dispersive cavity phase shift is due to a renormalization of the cavity center frequency by coupling to phonons \cite{Gullans17_PRB}. 

\begin{figure}[t]
	\centering
	\includegraphics[width=\columnwidth]{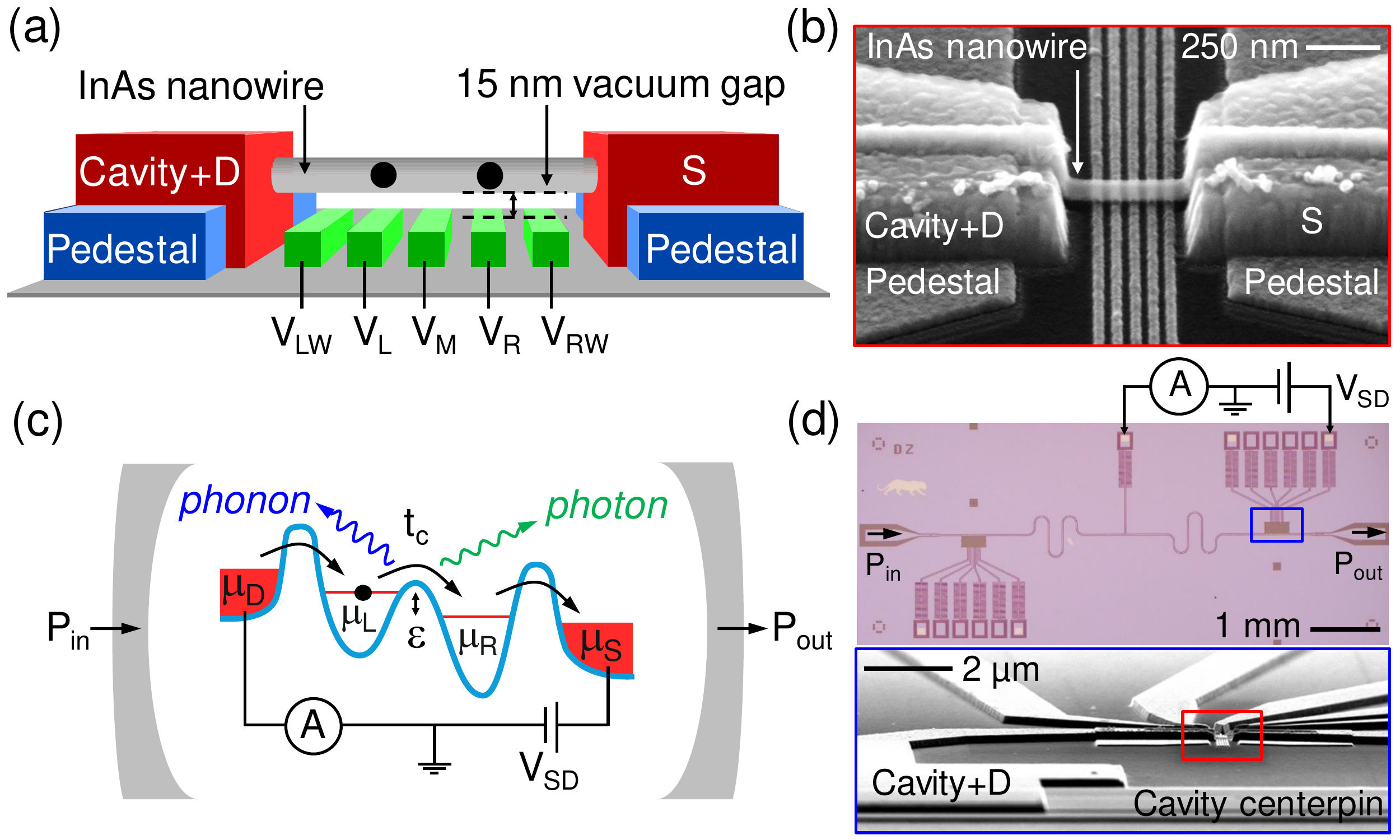}
	\caption{
	(a) Schematic representation of the suspended InAs nanowire DQD. 
	(b) Tilted angle SEM image of the device.
	(c) Energy level configuration of the DQD, which is placed inside a microwave cavity and probed by a weak microwave field. 
	(d) Optical image of the microwave cavity in (c) (upper panel), and tilted angle SEM image showing the electrical connection between the drain contact of the DQD and the cavity centerpin (lower panel).  The red box indicates the region shown in (b).}
	\label{fig:1}
\end{figure}

The suspended nanowire DQD device is shown schematically in Fig.~\ref{fig:1}(a). A 50 nm diameter InAs nanowire \cite{Schroer10_NanoLetters, Schroer10_RevSciInstrum} is manually placed across two lithographically defined Ti/Au pedestals using a long working distance optical microscope \cite{Flohr11_RevSciInstrum}. The pedestal thickness results in a 15 nm vacuum gap between the nanowire and Ti/Au electrostatic back gates (green) ($V_{\rm LW}$, $V_{\rm L}$, $V_{\rm M}$, $V_{\rm R}$, $V_{\rm RW}$) which are used to produce the double-well confinement potential. In contrast to earlier experiments \cite{NadjPerge10_Nature, Petersson12_Nature}, the back gates are not coated with SiN$_x$ dielectric. Instead, electrical isolation is achieved using the vacuum gap, which further confirms that the nanowire is physically suspended above the back gates. Ti/Au source (S) and drain (Cavity+D) contacts (red), with a separation of 380 nm, are deposited on top of the pedestals. A scanning electron microscope (SEM) image of a representative device is shown in Fig.~\ref{fig:1}(b). 

In order to study the interaction of electrons in the DQD with photons in the cavity and phonons in the nanowire, the DQD is electric-dipole coupled to the voltage antinode of a Nb superconducting transmission line resonator (Cavity+D) with resonance frequency $f_c = 7782.8$~MHz and quality factor $Q \sim 3050$ \cite{Liu15_Science, Liu17_PRL}. A schematic diagram illustrating the key elements of the device is shown in Fig.~\ref{fig:1}(c). The electrical connection between the DQD and cavity is shown in Fig.~\ref{fig:1}(d). Electron-phonon coupling is probed by measuring the dc current through the DQD and the cavity amplitude and phase response as a function of $\epsilon$ \cite{Stehlik15_PRA}. Measurements are performed in a dilution refrigerator with a base temperature $T$ = 10 mK.

\begin{figure}[t]
	\centering
	\includegraphics[width=\columnwidth]{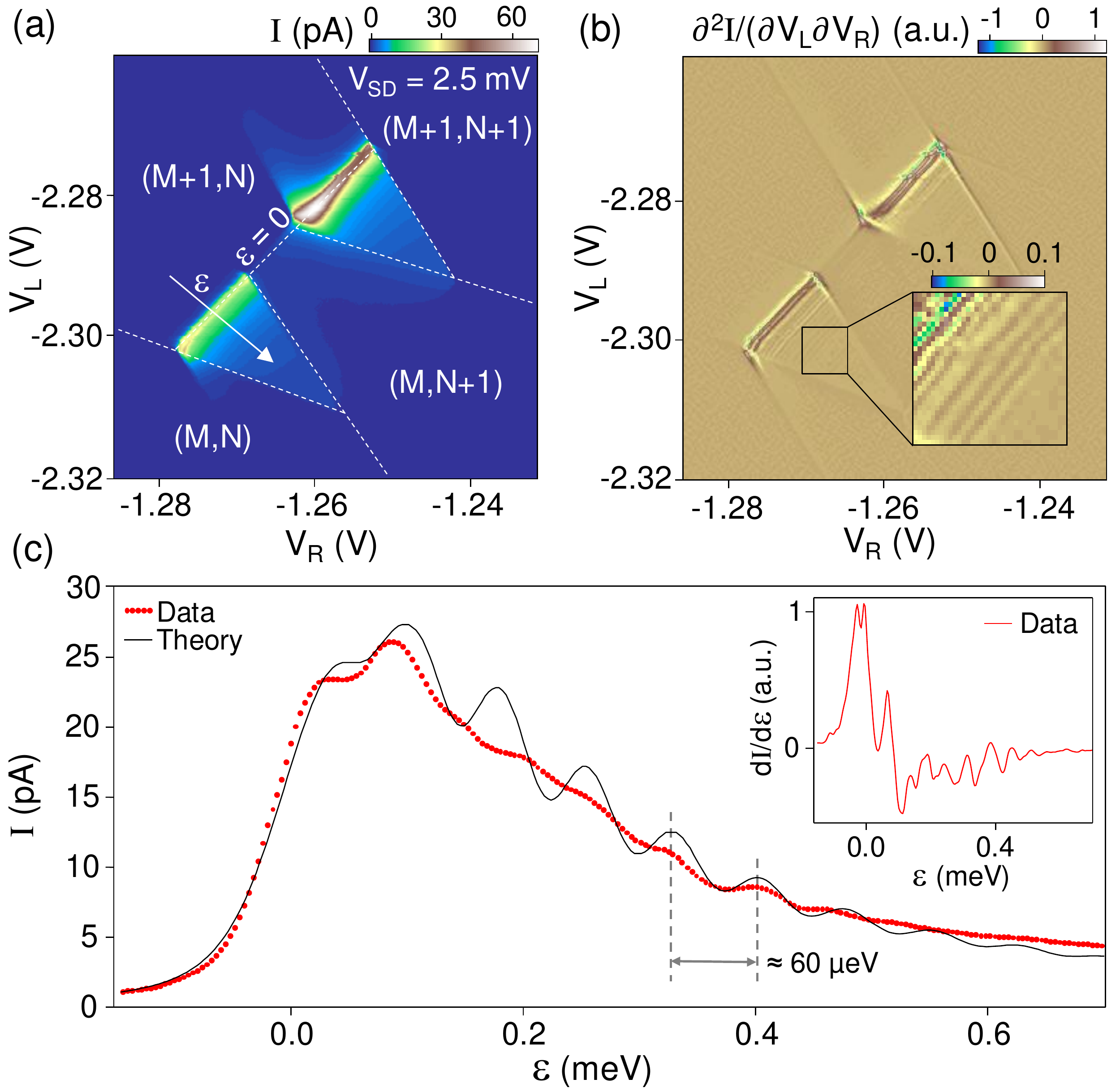}
	\caption{ 
	(a) Current $I$ through the DQD as a function of $V_{\rm{L}}$ and $V_{\rm{R}}$ with $V_{\rm{SD}}$ = $2.5$ mV. The interdot level detuning $\epsilon$ is adjusted along the white arrow.
	(b) Second derivative $\partial^2 I/(\partial V_{\rm{L}}\partial V_{\rm{R}} )$ of the data in (a) show periodic oscillations as a function of $\epsilon$.
	(c) $I$ as a function of $\epsilon$ (red) accompanied by a simple theory prediction (black) that accounts for phonon emission.  In the theory curve we use the parameters: $t_c = 22~\mu$eV, $J(2 t_c/\hbar)/2\pi =0.3$~GHz, $d=120~$nm, $\omega_0/2\pi = 130~$GHz,~$c_n=2\,100$~m/s, $r=4\times 10^{-3}$, $\Gamma_{L}/2\pi =\Gamma_{R}/2\pi=90$~MHz and a nanowire temperature of 200~mK. Inset: $dI/d\epsilon$ as a function of $\epsilon$ showing features with approximately 60 $\mu$eV periodicity in $\epsilon$. }
	\label{fig:2}
\end{figure}

Figure~\ref{fig:2}(a) shows the current $I$ through the device as a function of the gate voltages $V_L$ and $V_R$  for $V_{\rm SD}$ = 2.5 mV. Nonzero current is observed within finite bias triangles (FBTs), where the chemical potentials $\mu_D~>~\mu_L>\mu_R>~\mu_S$. DQD charge states are labeled by ($N_{\rm L}, N_{\rm R}$), where $N_{\rm{L}}$ ($N_{\rm{R}}$) is the number of electrons in the left (right) dot. We are able to reduce the electron number to the single electron regime, although the data presented in this paper are taken with $N_{\rm L} \approx N_{\rm R} \approx 4-6$ to increase the current through the device and therefore the number of photon emission events \cite{Liu14_PRL, Stockklauser15_PRL}. 

Early experiments on semiconductor DQDs showed that energy is conserved during inelastic interdot tunneling through spontaneous emission of a phonon \cite{Fujisawa98_Science}. Measurements of the current as a function of level detuning, $I(\epsilon)$, probe the environment at an energy $\Omega(\epsilon)~=~\sqrt{\epsilon^2 + 4 t_c^2}$, where $t_c$ is the interdot tunnel coupling. Oscillations in $I(\epsilon)$ were attributed to electron phonon coupling \cite{Fujisawa98_Science,Weber10_PRL, Roulleau11_NatCommun}. In contrast with these previous experiments, our DQD is electric-dipole coupled to a cavity. Therefore energy can be emitted during interdot tunneling by processes that involve the emission of both a phonon and a photon \cite{Liu14_PRL}, as illustrated in Fig.~\ref{fig:1}(c). Due to these higher-order emission processes, we may expect to see signatures of electron-phonon coupling in the amplitude and phase response of the cavity.

We first measure the current through the DQD and search for features in the data indicative of transitions between discrete electronic states that are accompanied by the emission of a phonon with energy $\Omega(\epsilon)$.  To increase the visibility of the current oscillations at this particular charge transition we plot $\partial^2 I/(\partial V_{\rm{L}}\partial V_{\rm{R}} )$ as a function of $V_{\rm{L}}$ and $V_{\rm{R}}$ in Fig.~\ref{fig:2}(b) over the same gate voltage range as Fig.~\ref{fig:2}(a). These data reveal features that are present throughout the inelastic region $(\epsilon \gg 0)$ of the FBT. The features are parallel to the interdot charge transition axis ($\epsilon = 0$) and therefore occur at a constant detuning, as expected for a decay mechanism that primarily depends on $\Omega(\epsilon)$. Figure~\ref{fig:2}(c) shows that small oscillations are directly visible in $I(\epsilon)$. A plot of $dI/d\epsilon$ (inset) shows that this current modulation has a periodic spacing of approximately 60 $\mu$eV. Current oscillations with a period of $60 \pm 5$ $\mu$eV were observed at more than 10 other charge transitions in the device, at forward and reverse bias (data not shown). 

To better understand the observed features in the current we model the interaction of the DQD with the lattice phonons using a spin-boson model \cite{Brandes99_PRL, Brandes05_PhysRep,Weber10_PRL}
\begin{equation}
H = \frac{\Omega(\epsilon)}{2} \sigma_z + \sum_k \hbar \omega_k a_k^\dagger a_k+\hbar \sum_k \lambda_k \sigma_x (a_k+a_k^\dagger),
\end{equation}
where $\sigma_z$ is the Pauli matrix acting on the two charge states of the DQD, $\omega_k$ is the dispersion of a phonon mode with index $k$, $\lambda_k$ is an electron-phonon interaction matrix element, and $a_k$ are bosonic operators for the phonon modes.  A key quantity in this model is the phonon spectral density $J(\nu) = \sum_k |\lambda_k|^2 \delta(\nu - \omega_k)$, which we approximate by the contribution from the lowest energy longitudinally polarized mode of the nanowire and a background term arising from other phonon modes in the system \cite{Gullans15_PRL}
\begin{align} \label{eqn:jnu}
J(\nu) &=  \frac{J_0 d}{c_n \nu}\sin^2( {\nu d}/{c_n} ) e^{-\nu^2/2\omega_0^2} + J_b(\nu), \\
J_b(\nu)&=r J_0 \frac{\nu c_n}{d} [1- \mathrm{sinc}(\nu d/c_n)] e^{-\nu^2/2\omega_0^2},
\end{align}
where $J_0$ is a constant scale factor, $c_n$ is the phonon speed of sound, $d$ is the spacing between the dots, $\omega_0$ is a cutoff frequency that scales with the size of each dot, and $r$ scales the relative contribution from the background.    From this idealized model, we see that the spectral density exhibits oscillations with the phonon energy $\hbar \nu$ when the DQD spacing $d$ is an integer multiple of the phonon wavelength $2\pi c_n/\nu$. This condition allows a simultaneous vibrational anti-node at the position of each DQD \cite{Brandes99_PRL}.  
Including the tunneling to the left (right) lead at rate $\Gamma_{L(R)}$ in the presence of a source-drain bias, we can derive analytic expressions for the current through the dot \cite{Brandes99_PRL,Gullans17_PRB}.  As seen in the theory curve in Fig.~2(c), modulations in the phonon spectral density show up directly in the current because the charge relaxation rate of the DQD two-level system is directly proportional to $J[\Omega(\epsilon)/\hbar]$. 

\begin{figure}[t]
	\centering
	\includegraphics[width=\columnwidth]{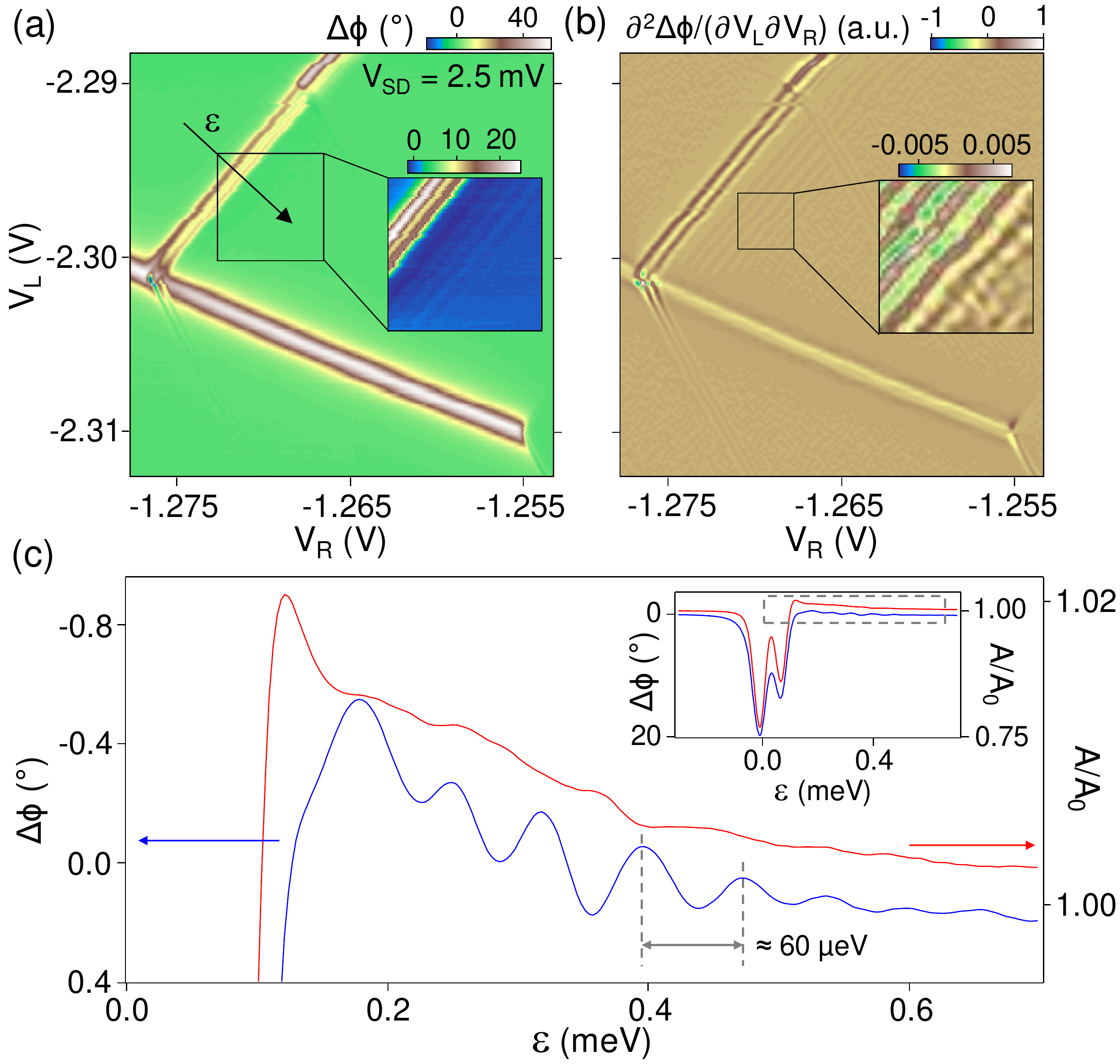}
	\caption{(a) Cavity phase response $\Delta \phi$ as a function of $V_{\rm{L}}$ and $V_{\rm{R}}$ near the lower FBT. 
	(b) Second derivative $\partial^2 \Delta \phi/(\partial V_{\rm{L}}\partial V_{\rm{R}} )$ of the data in (a).	
	(c) $\Delta \phi$ and $A/A_0$ as a function of detuning $\epsilon$. Inset: Full range of data. The subset of data shown in the main panel is outlined by the dashed grey box.  $A/A_0$ is normalized to the value in Coulomb blockade.}
	\label{fig:3}
\end{figure}

Going beyond previous experiments, we now report the observation of features in the amplitude and phase response of the cavity that have the same 60 $\mu$eV periodicity as the features observed in the current. The cavity transmission, $A/A_0$, and phase response, $\Delta \phi$, are investigated in Fig.~\ref{fig:3} for the lower FBT of the same interdot transition shown in Fig.~\ref{fig:2}. These measurements are performed by driving the cavity with an input tone at $f_c = 7782.8$ MHz resulting in an intra-cavity photon number $n_c \approx 50$. The phase response of the cavity [Fig.~\ref{fig:3}(a)] exhibits features that are once again periodic as a function of detuning (inset), and present throughout the FBT, suggesting they are caused by an energy-dependent decay mechanism. These features are more clearly visible in $\partial^2 \Delta \phi/(\partial V_{\rm{L}}\partial V_{\rm{R}} )$ [Fig.~\ref{fig:3}(b)]. Figure~\ref{fig:3}(c) shows $A/A_0$ and $\Delta \phi$ as a function of detuning $\epsilon$ in the lower FBT, measured along the same detuning axis as in Fig.~\ref{fig:2}(c). At positive detuning, clear features are visible in $\Delta \phi$ with a period of approximately $ 60$ $\mu$eV. Data are shown over a larger range of detuning in the inset. The dip in transmission and large phase shift near $\epsilon = 0$ are due to the dispersive interaction of the DQD and cavity photons \cite{Petersson12_Nature}. The second dip in transmission around $\epsilon \approx 70$ $\mu$eV may be due to a low lying excited state, although the energy scale is more consistent with the periodic features observed in $I$, $A/A_0$, and $\Delta \phi$. The features that we observe at large detuning in $A/A_0$ and $\Delta \phi$ are also independent of small perturbations in the barrier gate voltage, $V_M$, suggesting they are a robust consequence of coupling to the phonon bath.
 
\begin{figure}[t]
	\centering
	\includegraphics[width = \columnwidth]{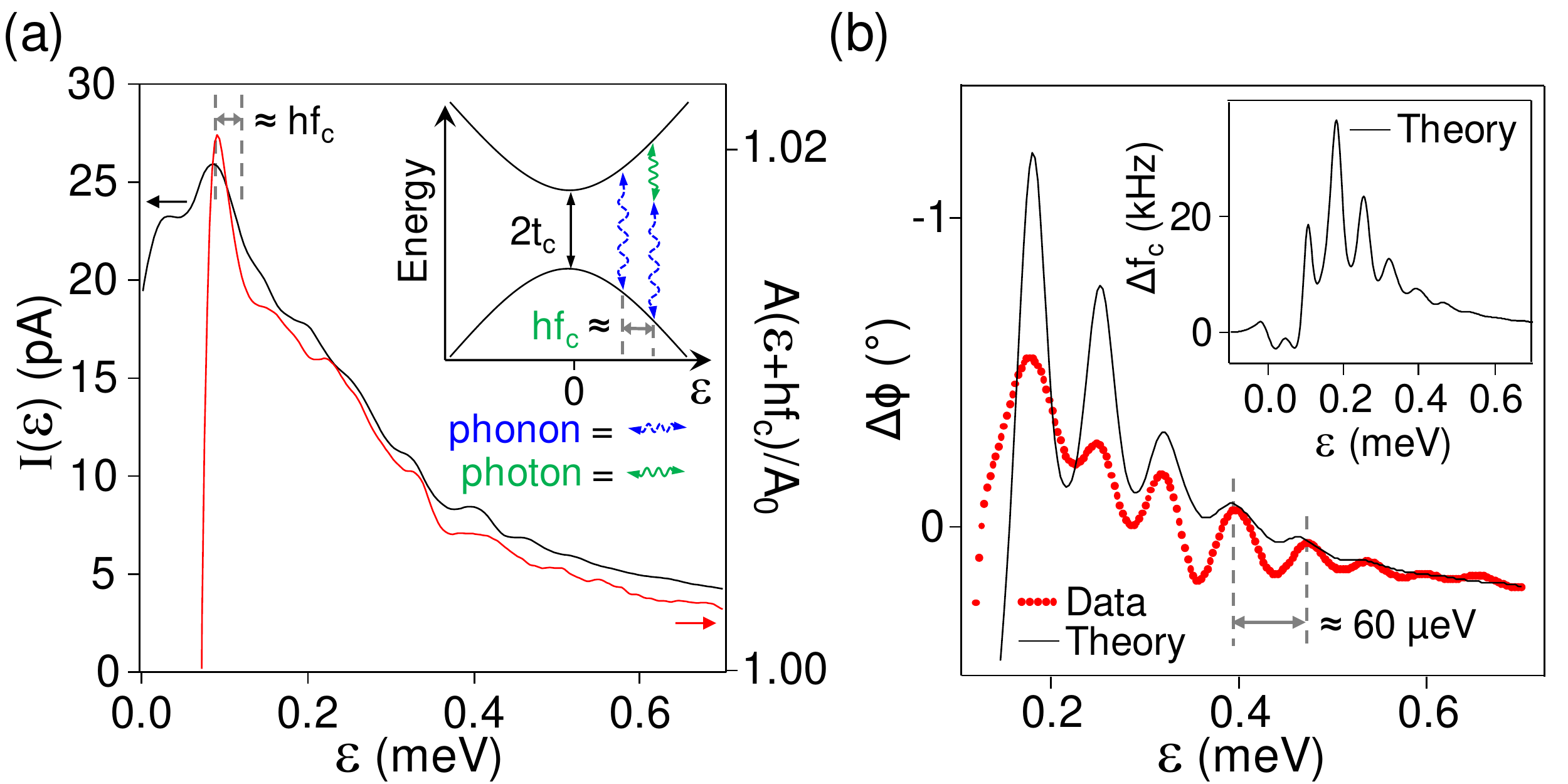}
	\caption{
	(a) A comparison of $I(\epsilon)$ and $A(\epsilon + hf_c)/A_0$.  Inset: DQD energy level diagram showing the first order emission of a phonon (blue) or the second order emission of the same phonon and a photon (green) of energy $hf_c$, which occurs when $\epsilon$ increases by approximately $hf_c$.
	(b)	 $\Delta \phi$ as a function of $\epsilon$ (red) and theory fit (black). The theory curve takes into account the dispersive cavity shift (inset) that is caused by the coupling of cavity photons to nanowire phonon modes. These data, and the Fig.~\ref{fig:2}(c) data, are simultaneously fit.}
	\label{fig:4}
\end{figure}

It is helpful to search for correlations between the oscillations observed in the current and the oscillations observed in the cavity response. In general, the energy emitted into the environment during inelastic interdot tunneling is given by the DQD energy splitting $\Omega(\epsilon)$. Energy can be released into the environment by creating phonons and cavity photons. Previous experiments show that roughly one photon is emitted into the cavity mode for every 10$^3$--10$^4$ electron tunneling events \cite{Liu14_PRL}. Therefore the electronic current primarily probes the phonon environment and should scale as $I(\epsilon)/|e| \propto J[\Omega(\epsilon)/\hbar]$. In contrast, second order processes, where a cavity photon and phonon are emitted during inelastic tunneling, scale with the phonon spectral density $J(\nu)$ as $J[\Omega(\epsilon)/\hbar - 2\pi f_c]$. These processes are illustrated in the inset of Fig.~\ref{fig:4}(a). For the large level detunings examined here, $|\epsilon| \gg t_c$, and $\Omega \approx \epsilon$. As such, there should be a correlation between the measured current $I(\epsilon)$ and cavity response $A(\epsilon + hf_c)/A_0$.

Figure~\ref{fig:4}(a) plots $I(\epsilon)$ and $A(\epsilon + hf_c)/A_0$. The y-axes offsets and data ranges have been adjusted such as to maximize overlap between the two distinct data sets. For $\epsilon$ $>$ 0.1 meV there is a strong correlation between $I(\epsilon)$ and $A(\epsilon + hf_c)/A_0$. Both data sets have a very similar inelastic tail. Moreover, the oscillations in 
$I(\epsilon)$ occur at the same values of detuning as the oscillations in $A(\epsilon + hf_c)/A_0$. These data give further evidence that the oscillations in the current and cavity response are due to the same electron-phonon coupling mechanism.

The cavity phase response can be modeled using a theory that takes into account the dispersive shift of the cavity (renormalization of the cavity frequency) due to the DQD-mediated coupling to nanowire phonons \cite{Gullans17_PRB}. Measurements of $\Delta \phi$ and best fits to the microscopic theory are shown in Fig.~\ref{fig:4}(b). The theoretical predictions for the dispersive cavity shift due to electron-phonon coupling are shown in the inset of Fig.~\ref{fig:4}(b).  In comparison with previous theoretical analysis of these systems \cite{Gullans15_PRL,Muller17_PRA}, our theory accounts for  small energy shifts in the cavity frequency that arise from resonant coupling of the cavity photons to the nanowire phonons when the DQD is in the excited state \cite{Gullans17_PRB}.  The periodic oscillations in the phase arise because  periodic modulations in the charge relaxation rate lead to a similar modulation in the excited state population, which shifts the cavity frequency, as seen in the inset of Fig.\ 4(b).  The microscopic origin of the oscillations in our model is identical across all three independent measurement techniques of dc current $I$, cavity amplitude $A/A_0$, and phase $\Delta \phi$ [see Eq.~(\ref{eqn:jnu})], which provides strong evidence that we have observed direct signatures of the electron-phonon coupling in this system. 

Future work on suspended nanowires could explore the dependence of the phonon periodicity on the source-drain electrode spacing. Such experiments could help to resolve the conflicting interpretations of the cause of this periodicity in Refs.~\cite{Weber10_PRL} and \cite{Roulleau11_NatCommun}. More broadly, the principles of phonon spectrum engineering and measurement suggested by this work may help to minimize electron-phonon interaction processes at specific energies. For example, single spin relaxation times in III/V semiconductor quantum dots  \cite{Elzerman04_Nature, Meunier07_PRL} are due to spin-orbit coupling and phonon emission \cite{Khaetskii00_PRB, Khaetskii01_PRB, Hanson07_RevModPhys}. By tailoring the phonon spectrum, it may be feasible to extend electron spin lifetimes in quantum devices \cite{Trif08_PRB}.

In summary, we have shown that it is possible to create a cavity-coupled InAs nanowire DQD that is mechanically suspended above the substrate. Consistent with earlier work, we observe oscillations in the inelastic current as a function of level detuning due to electron-phonon couping in the nanowire \cite{Weber10_PRL, Roulleau11_NatCommun}. Measurements of the cavity response are also sensitive to electron-phonon coupling. We couple the electronic dipole moment of an electron trapped in this DQD to the electric field of a microwave cavity and observe a periodic cavity phase response due to a dispersive interaction with nanowire phonons. A comparison of these measurements with a microscopic theoretical model of the device suggests that the coupling of phonons to photons, mediated by electron dynamics, results in a phonon renormalization of the cavity center frequency. These experiments broadly help to understand the fundamental nature of electron-phonon coupling in nanoscale systems and may provide a path toward mitigating spin decay in semiconductor quantum devices. 

\begin{acknowledgements} 
Supported by the Packard Foundation, the Gordon and Betty Moore Foundation's EPiQS Initiative through grant GBMF4535, and NSF grants DMR-1409556 and DMR-1420541.  Devices were fabricated in the Princeton University Quantum Device Nanofabrication Laboratory.
\end{acknowledgements}

\bibliographystyle{apsrev_lyy2017} 
\bibliography{Hartke_Suspended_PRL_2017_references_v3}

\newpage

\end{document}